\documentclass[prl,twocolumn,showpacs,amsmath,amssymb]{revtex4}

\usepackage[dvips]{graphicx}

\usepackage{bm}

\begin{document}

\title{Anomalous Hall effects and electron polarizability}

\author{Pavel St\v{r}eda$^{1}$ and Thibaut Jonckheere$^{2}$}

\affiliation{$^{1}$Institute of Physics, Academy of Sciences of the
Czech Republic, Cukrovarnick\'{a} 10, 162 53 Praha, Czech Republic}
\affiliation{$^{2}$Centre de Physique Th\'{e}orique, Facult\'e de 
Luminy, Case 907, 13288 Marseille, France}

\date{\today}

\begin{abstract}

A theory of the anomalous and spin Hall effects,
based on the space distribution of the current densities,
is presented. Spin-orbit coupling gives rise
to a space separation of the mass centers,
as well as a current density separation of the
quasiparticle states having opposite group velocities.
It is shown that this microscopic property is essential
for existence of both Hall effects.
\end{abstract}


\maketitle

It has been known for more than a century that a ferromagnetic material
exhibits, in addition to the standard Hall effect when placed in a magnetic
field, an extraordinary Hall effect which does not vanish at
zero magnetic field. The theory of this so-called anomalous Hall effect
(AHE) has a long and confusing history, with different approaches
giving in some cases conflicting results. 
While more recent calculations have somewhat unified the different
approaches and clarified the situation, it is still an active topic
of research (see Ref. [\onlinecite{Nagaosa2009}] for a
recent review).
Closely related to the AHE is the spin Hall effect,
represented by spin accumulation on the edges of a current
carrying sample \cite{spin_Hall_1,spin_Hall_2}.

It is generally accepted that the anomalous and spin Hall effects
are induced by spin-orbit coupling. It was first suggested
by Karplus and Luttinger \cite{Karplus} in 1954 to explain
anomalous Hall effect observed on ferromagnetic crystals.
Their analysis leads to the scattering independent
off-diagonal components of the conductivity,
which are assigned to the so-called "intrinsic" effect.
Later, theories of this effect based on several specific 
models have been developed \cite{Miyazawa,Onoda}.
As has been recently shown it is accompanied by
strong orbital Hall effect \cite{Shindou,Kontani}. 
The conductivity is also affected by the scattering,
which in the presence of spin-orbit coupling gives rise
to so called side-jump \cite{Berger} and skew scattering
\cite{Smit,Luttinger_58,Sinitsyn}. These also
lead to anomalous Hall effect, called "extrinsic".
The best quantitative agreement with experimental observations
has been obtained with semi-classical transport theory
\cite{Niu}, leading to the Berry phase correction to the group
velocity. For Fe crystals \cite{Jungw_2} it gives an anomalous
conductivity $\sim 750 \; \Omega^{-1}$ cm$^{-1}$  while
a value approaching $ 1000 \; \Omega^{-1}$ cm$^{-1}$
has been observed.

It is the goal of this letter to shed light on AHE using a novel
point of view. Our approach is based on the analysis of the 
space distribution of local current densities, and it is simple and
rather intuitive. We will show that the anomalous Hall conductivity
is related to the spatial separation of the mass centers of states
with opposite velocities.
This confirms the interpretation of AHE in ferromagnetic
systems as a consequence of the periodic field of electric dipoles
(electric polarizability) induced by the applied
current~\cite{Karplus,Adams,Fivaz}, despite the fact that the
original arguments were not convincing~\cite{Berger}.
We will also show that in non-magnetic systems, the spin-orbit coupling
leads to a periodic variation of the spin polarizability
of the current densities in the transport regime. This effect can
be viewed as an internal spin Hall effect.
For the sake of simplicity, we limit our consideration
to crystalline structures invariant under
space inversion.

Current density distribution is closely related to the
orbital magnetization of solids. In zero magnetic
field it has its origin in the orbital
magnetic moment of atomic states. Ignoring spin effects,
atomic wave functions in spherical polar-coordinate system
can be written as
\begin{equation}
\Psi^{(at)}_{\alpha}(r,\theta,\phi) \, = \,
f_{\alpha}(r,\theta) \, \frac{e^{im \phi}}{\sqrt{2 \pi}}
\qquad , \quad \alpha \, \equiv \, n , l , m 
\; ,
\end{equation}
where $m$ is the so called magnetic quantum number,
with $m= 0, \pm 1, \cdots$, and $|m| \le l$. It 
determines the magnetic moment along the $\hat{z}$ direction
\begin{equation} \label{M_z(m)}
M_z(\alpha) \equiv -\frac{e}{2 c}
\langle \alpha | \left( \vec{r} \times \vec{v} \right)_z
| \alpha \rangle\equiv
-\frac{m}{|m|} \, \frac{\pi R_{\alpha}^2}{c} \; \left|j_{\alpha}\right|
\; ,
\end{equation}
where $e$ denotes absolute value of the electron charge and $\vec{v}$
stands for velocity operator.
The last expression represents a classical analogy with
$j_{\alpha}$ being the current flowing on a circular
loop of the radius $R_{\alpha}$. 
Because of the energy degeneracy in $m$ the total orbital
magnetic moment vanishes. However, spin-orbit coupling
together with exchange interaction remove this
degeneracy giving rise to non-zero magnetic moment.

Within a mean field approach the electron properties are controlled
by a single electron Hamiltonian $H$ containing two additive
terms $H_{so}$ and $H_{z}$ representing spin-orbit coupling
and an effective Zeeman-like spin splitting due to the exchange 
interaction, respectively:
\begin{equation}
H \, = \, \frac{p^2}{2m_0} \, + \, V(\vec{r}) \, + \,
H_{so} \, + \, H_{z}
\; ,
\end{equation}
with $m_0$ being free electron mass, $V(\vec{r})$ denotes the 
crystalline potential, $\vec{p}$
is momentum operator and
\begin{equation}
H_{so} \, = \, \frac{\lambda_c^2}{4 \hbar} \,
\vec{\sigma} \cdot
\left[ \vec{\nabla} V(\vec{r}) \times \vec{p} \, \right]
\; \; , \; \;
H_{z} \, = \, - \, \mu_B \, \vec{B}_{\rm eff} \cdot \vec{\sigma}
\, , 
\end{equation}
where $\lambda_c$ denotes an effective Compton length,
and elements of the vector $\vec{\sigma}$ are Pauli matrices.
Strength of the Zeeman-like splitting is controlled by the product 
of the Bohr magneton $\mu_B$ and the parameter $\vec{B}_{\rm eff}$
representing an effective magnetic field.
The corresponding velocity operator reads
\begin{equation} \label{v_def}
\vec{v} \, = \, \frac{\vec{p}}{m_0} \, + \,
\frac{\lambda_c^2}{4 \hbar} \,
\vec{\sigma} \times \vec{\nabla} V(\vec{r})
\; .
\end{equation}
Eigenfunctions are spinors with two components,
and since spin-orbit coupling does not destroy translation
symmetry they are of the Bloch form.
Energy spectrum $E_{\alpha}(k)$ is a
function of the wave vector $\vec{k}$, with $\alpha$ being a
band index, now including also, in addition to atomic orbital numbers,
a spin number. Eigenfunctions
are of the following form
\begin{equation}
| \alpha , \vec{k} \rangle  \equiv 
\Psi_{\alpha,\vec{k}}(\vec{r}) =
\frac{e^{i \vec{k} \vec{r}}}{\sqrt{8 \pi^3}} \,
u_{\alpha}(\vec{k},\vec{r})
\, ,
\end{equation}
and velocity expectation values are
\begin{equation} \label{v_expect}
\vec{v}_{\alpha}(\vec{k}) = \frac{1}{\hbar}
\vec{\nabla}_{\vec{k}} E_{\alpha}(\vec{k})
\, .
\end{equation}
Spinors $u_{\alpha}(k,\vec{r})$ are periodic
functions of the lattice translation vectors.
Assumed invariance under space inversion results
in following $\vec{k}$-space symmetry:
$E_{\alpha}(\vec{k}) = E_{\alpha}(-\vec{k})$ and
$v_{\alpha}(\vec{k}) = -v_{\alpha}(-\vec{k})$.

In order to analyze the role of the space distribution of the current
densities, it is illustrative to present first the results
for a simple model of a linear chain of atomic orbitals.
It is assumed that this chain is forming a one-dimensional
lattice along $\hat{x}$ direction with a period $a$.
Model parameter will be chosen to satisfy conditions
for which the tight-binding approach is applicable.
Energy bands originated in overlap of atomic states $|\alpha \rangle$
will be denoted by the corresponding magnetic quantum number $m$.
To model a ferromagnetic state, we assume that the
effective field $\vec{B}_{\rm eff}$ is parallel with
$\hat{z}$ direction and it is strong enough to
ensure spin orientation along $\pm \hat{z}$ direction
with $s_z = \pm 1/2$ being good quantum numbers.
We have obtained numerical results by diagonalizing the
single particle Hamiltonian for
a two-dimensional separable chain potential
\begin{equation} \label{model_pot}
V(x,y)\, = \, - \, V_0 \cos(2 \pi x /a) + m_0 \Omega_0^2 y^2 /2
\; .
\end{equation}
Note that adding a $z$-dependent potential would not affect the
current distribution of the considered model system.
The parameters have been chosen to be in the tight-binding regime,
i.e. to fully separate the studied band from the other energy bands
(we used $2 m_0 a^2 V_0/\hbar^2 = 75.0$,
$m_0 \Omega_0^2 a^2 = 1.4\times 4 \pi^2 V_0$ 
and $\pi \lambda_c^2/a^2 = 0.015$).

\begin{figure}[h]
\includegraphics[angle=0,width=3.3 in]{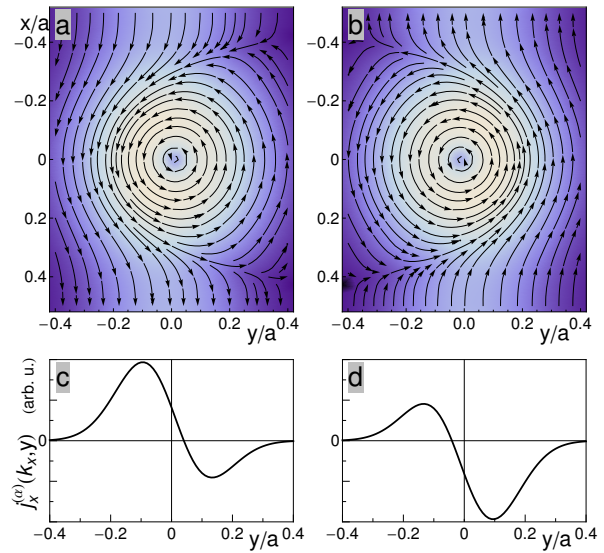}
\caption{(color online) Current distributions for an energy band
given by a chain of atomic states with spin $s_z=1/2$ and $m=-1$ for
$k_x= 1.5/a$ (a,c) and $k_x= - 1.5/a$ (b,d). In a and b,
the arrows indicate the direction of the current,
and a lighter background indicates a larger current.
In c and d the averaged current densities $j_x^{\alpha}(k_x,y)$
are shown.}
\label{j_distribution}
\end{figure}

Typical current density distribution within the 
unit cell for the Bloch states $|\alpha , k_x \rangle$
and $| \alpha , -k_x \rangle$ are shown in Figs.~1a and 1b,
respectively. One observes circulating currents forming vortices,
which have the same orientation for both cases. This orientation
coincides with the orientation of the circulating current of the atomic
orbitals. In addition to this circulating current, there is a
direct current flow, with opposite sign for the two cases.
Corresponding total current and its direction is just
determined by the velocity expectation value, $- e v_x^{\alpha}(k_x)$.
Note that these current flows are spatially separated in the
two cases (they are on opposite sides of the circulating
current).  The current densities averaged over
$x$ and $z$ coordinates,
\begin{equation} \label{j_x(k,y)}
j_x^{\alpha}(k_x,y) = - e \int \hspace{-2mm} \int
\Psi_{\alpha,k_x}^+(\vec{r}) \, v_x \Psi_{\alpha,k_x}(\vec{r}) \, dx \, 
dz
\, ,
\end{equation}
clearly demonstrate the above mentioned spatial separation
of the currents having opposite velocity directions,
as shown on Figs.~1c and 1d.

This space separation of currents flowing in
opposite directions is closely related to the mass-center separation
$\Delta Y_{\alpha}(k_x)$ of states $|\alpha,k_x\rangle$ and
$|\alpha,-k_x\rangle$ with $k_x$ being consider as positive,
$k_x>0$.
For the considered model potential, Eq.~(\ref{model_pot}),
the $y$-component of the force operator reads
\begin{equation}
F_y \, \equiv \, \frac{1}{i \hbar} \, [p_y , H ] \, = \, - \,
m_0 \Omega_0^2 y \, + \, 2 s_z \,
\frac{\lambda_c^2}{4 \hbar}  \, m_0 \Omega_0^2 \, p_x
\; .
\end{equation}
In a stationary state the force expectation values has to vanish.
Using the relation between $p_x$ and $v_x$ given by Eq.~(\ref{v_def})
we get
\begin{equation} \label{Delta_Y_model}
\Delta Y_{\alpha}(k_x) =
2 s_z \, \frac{\lambda_c^2 m_0}{4 \hbar}
\left[ 1 - \left(
\frac{\lambda_c^2 m_0}{4 \hbar} \right)^2 \Omega_0^2 \right]^{-1}
\! 2 v_{x}^{\alpha}(k_x)
\; . 
\end{equation}
In the limiting case of vanishing spin-orbit coupling
$\Delta Y_{\alpha}(k_x) \rightarrow 0$ and sum of the current
densities $j_x^{\alpha}(k_x,y)+j_x^{\alpha}(-k_x,y)$
approaches zero as well.

Non-zero total current appears if there is different occupation
of states with opposite velocities which can be characterized
by the chemical potential difference $\Delta \mu$. It can be
related to the electric field along $\hat{y}$ direction,
${\cal{E}}_y = \Delta \mu /(e |\Delta Y_{\alpha}(k_F)|)$,
with $\Delta Y_{\alpha}(k_F)$ being the mass-center separation of
quasiparticles having opposite velocities at the Fermi energy
$E_F=\mu$. Within linear response approach the resulting current
at zero temperature reads
\begin{equation} \label{current}
J_x^{\alpha}(\mu) = - \frac{e}{h}
\frac{v_{x}^{\alpha}(k_F) - v_{x}^{\alpha}(-k_F)}
{\left| v_{x}^{\alpha}(k_F) \right|} \,
\Delta \mu = - 
\frac{e^2}{h} \, \Delta Y_{\alpha}(k_F) \, {\cal{E}}_y 
\; .
\end{equation}
Because of the non-zero separation $\Delta Y_{\alpha}(k_F)$
and non-equal occupation of states with opposite velocities
the applied current is giving rise an electric dipole moment,
i.e. a charge polarization is induced.

For the later use, let us express current $J_x^{\alpha}(\mu)$
in terms of the following quantity
\begin{equation} \label{OPM}
\frac{\vec{P}_{\alpha}(\vec{k})}{\Omega_{\rm ws}} \, \equiv \, - \,
\frac{e}{c} \,
\langle \alpha, \vec{k} | \vec{r} | \alpha, \vec{k} \rangle 
\times \vec{v}_{\alpha}(\vec{k})
\; ,
\end{equation}
where $\Omega_{\rm ws}$ defines a unit cell volume.
Evaluation of the following expression 
\begin{equation}
J_x^{\alpha}(\mu) = - \frac{ec {\cal{E}}_y}{2 \pi \Omega_{\rm ws}}
\int\limits_{-\pi/a}^{+\pi/a} \! \!
\delta(E_{\alpha}(k_x) - \mu) \,
\left[ \vec{P}_{\alpha}(k_x) \right]_z dk_x
\; ,
\end{equation}
immediately gives the above result, Eq.~(\ref{current}), since
velocity expectation value along $\hat{y}$ direction vanishes.
The above defined quantity, Eq.~(\ref{OPM}), is the part of
the orbital magnetic moment within each of the unit
cells which gives rise an electric dipole moment in the current
carrying regime. For this reason it will be called
as the orbital polarization moment.

Generalization of the above treatment to a three-dimensional
system is straightforward.
Velocity expectation values have non-zero component also
along $\hat{y}$ direction and they contribute to the orbital 
polarization moment defined by Eq.~(\ref{OPM}). The resulting
contribution of the band $\alpha$ to the Hall conductivity
component $\sigma_{xy}$ can thus be
written as follows 
\begin{equation} \label{sigma_OPM}
\sigma_{xy}^{(\alpha)}(\mu) = -
\frac{ec}{8 \pi^3 \Omega_{\rm ws}}
\int\limits_{\rm BZ} \! \! 
\delta(E_{\alpha}(\vec{k}) - \mu) \,
\left[ \vec{P}_{\alpha}(\vec{k}) \right]_z d^3k
\, ,
\end{equation}
where  integration is limited to the Brillouin zone
and $\Omega_{\rm ws}$ now denotes volume of the
Wiegner-Seitz cell. Inserting for $\vec{P}_{\alpha}(\vec{k})$
and $\vec{v}_{\alpha}(\vec{k})$ their explicit forms,
Eq.~(\ref{OPM}) and Eq.~(\ref{v_expect}), respectively,
and using equality
\begin{equation}
\langle \alpha, \vec{k} | \vec{r} \, |\alpha, \vec{k} \rangle = - 
{\rm Im} \int\limits_{\Omega_{\rm ws}} \!
u^{+}_{\alpha}(\vec{k},\vec{r}) \left( \vec{\nabla}_{\vec{k}}
u_{\alpha}(\vec{k},\vec{r}) \right) d^3r
\, ,
\end{equation}
already derived by Karplus and Luttinger \cite{Karplus},
the integration per parts gives the well known expression
for the Hall conductivity of Bloch electrons
\begin{equation} \label{sigma_Berry}
\sigma_{xy}(\mu)  = - \frac{e^2}{4 \pi^2h}
\sum_{\alpha} \int\limits_{\rm BZ} \! \!
f_0\left( E_{\alpha}(\vec{k}) \right) \,
\left[ \vec{\Omega}_{\alpha}(\vec{k}) \right]_z d^3 k
\, .
\end{equation}
Here $f_0(E)$ stands for zero-temperature Fermi-Dirac
distribution function and the Berry phase curvature
$\vec{\Omega}_{\alpha}(\vec{k})$ defined by the periodic
part of Bloch functions, $u_{\alpha}(\vec{k},\vec{r})$,
reads
\begin{equation} \label{Berry}
\vec{\Omega}_{\alpha}(\vec{k}) \, = \, - \, {\rm Im} \,
\left\langle \, \vec{\nabla}_{\vec{k}} \, u_{\alpha}(\vec{k}, \vec{r})
\, \left| \times \right| \,
\vec{\nabla}_{\vec{k}} \, u_{\alpha}(\vec{k}, \vec{r}) \, \right\rangle
\, .
\end{equation}
Our description of the anomalous Hall effect based on charge 
polarization effect is thus equivalent to the approach based on the
Berry phase correction \cite{Niu}. 

For completeness, we must note that for the three-dimensional case,
computing the quantity $\Delta Y$, representing the mass-center
separation of states having opposite velocities along
$\hat{x}$ direction, is not as easy as it was for
the single atomic chain. It requires to express eigenfunctions
in a mixed representation, to preserve the Bloch form
along the $\hat{x}$ direction, while using Wannier representation
along perpendicular directions. The Wannier representation gives
functions which are bounded along the $\hat{y}$ direction, allowing
to compute $\Delta Y$. Although
the explicit form of  $\Delta Y$ is not simple, the main features
are qualitatively the same as those presented for the
single chain. 

Conductivity can directly be measured on samples having
Corbino disc geometry. This arrangement can be modelled by
considering a strip with periodic boundary conditions
along $\hat{x}$ direction and current contacts attached
to the strip edges allowing to apply current along
$\hat{y}$ direction. It induces gradient of the electro-chemical
potential along $\hat{y}$ direction and consequently
it gives rise to the local charge polarization. As a result, 
Hall current is induced. Intra band scattering ensuring
finite conductivity across the strip is naturally
of the side-jump character. This type of scattering
does not affect Hall conductivity which is given as the
sum of the additive band contributions defined by
Eq.~(\ref{sigma_OPM}) or Eq.~(\ref{sigma_Berry}).
However, inter band scattering which is generally
of the skew character can affect Hall conductivity
significantly. These features of the anomalous conductivity
directly follows from the analysis based on quantum
theory of the linear response, Kubo formula.
Although the derivation is straightforward, it is quite
lengthy and will be thus presented in a separate publication.

The relation we have presented here between the charge polarization and
the anomalous Hall effect is similar to the discussion of Hall
conductivity of Bloch electrons in rational quantizing magneticgv 
fields in terms of charge polarization.\cite{polarizability_08}
However, in that case the physical picture is strongly affected
by chiral magnetic edge states. 

\begin{figure}[h]
\includegraphics[angle=0,width=3.3 in]{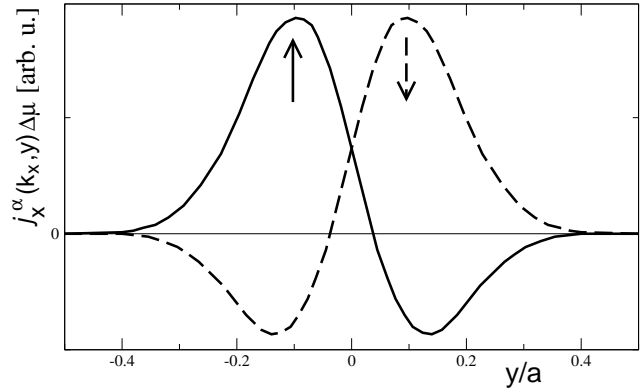}
\caption{Transport current densities $j_x^{\alpha}(k_x,y) \Delta \mu$
for the energy band of the two-fold degeneracy ($B_{\rm eff}=0$)
given by a chain of atomic states $|\alpha \rangle$
with $m=-1, \, s_z=1/2$ (full line) and  $m=1, \, s_z=-1/2$
(dashed line) for Fermi energy given by the wave number
$k_x = 1.5/a$.}
\label{j_spin}
\end{figure}

Of particular interest are non-magnetic systems in which
spin-orbit coupling is not negligible but effective Zeeman-like
spin splitting vanishes, $B_{\rm eff} \rightarrow 0$.
In this case the states of the single atomic chain
$|\alpha , k_x \rangle$ with orbital number $m$ and
spin $s_z$ are of the same energy as states
$|\bar{\alpha} , k_x \rangle$ with  opposite sign
of the orbital number and spin, $-m$ and $-s_z$.
Their orbital magnetization have opposite sign,
the sum of their current densities vanishes,
$j_x^{\alpha}(k_x,y) + j_x^{\bar{\alpha}}(k_x,y) +
j_x^{\alpha}(-k_x,y) + j_x^{\bar{\alpha}}(-k_x,y) = 0$, and
total magnetization vanishes as well.
The mass-center separation has also opposite signs, 
$\Delta Y_{\alpha}(k_x) = - \Delta Y_{\bar{\alpha}}(k_x)$,
and in accord with Eq.~(\ref{current}) the resulting
anomalous Hall effect vanishes.
However, the spin-orbit coupling has still an important effect
in the transport regime. The current applied along the $\hat{x}$
direction gives rise for each band to non-equal occupation of states with
opposite velocities represented by a local chemical potential
difference $\Delta \mu$. The two considered bands have the same total current,
but they have different space distribution because of the different
mass-center positions determined by their spin orientation
(Eq.~(\ref{Delta_Y_model})). As a result, the spin polarization
of the transport current density averaged over $x$ and $z$
coordinates will be a function of the $y$ coordinate.
This is illustrated on Fig.~\ref{j_spin}, where the averaged
transport current densities are shown for the same model parameters
as in Fig.~\ref{j_distribution}. 
Qualitatively the same features are expected for 
three-dimensional crystals: the spin polarization of the
transport current density will show a periodic oscillation.
This property can be interpreted as an internal spin Hall
effect. At the sample edges of semiconductor systems,
the oscillations of the spin polarizability will be modified
by the confining potential defining sample edges.
It can be expected that this modification is responsible
for the already observed spin Hall effect.
\cite{spin_Hall_1,spin_Hall_2}

To conclude, we have shown that the mass-center separation
as well as the current density separation of states having
opposite velocities is the essential feature of the
systems with spin-orbit coupling. In the transport regime
it gives rise to the charge polarization inducing
anomalous Hall effect in ferromagnetic crystals.
In non-magnetic systems it leads to a periodic
spatial variation of the spin polarizability of the
transport current density,
predicted internal spin Hall effect,
which is expected to be the origin of the spin accumulation
at the edges of current currying samples.

Authors acknowledge support from
Grant No. GACR 202/08/0551
and the Institutional Research Plan No. AV0Z10100521.
P.S. thanks CPT (UMR 6207 of CNRS) and Universit\'e du Sud
Toulon-Var for their hospitality.

\end{document}